\documentclass[12pt]{iopart}
\usepackage{etoolbox}
\csundef{equation*}
\csundef{endequation*}
 
\usepackage{graphicx}
\usepackage{amsmath}
\usepackage{amssymb}
\usepackage{epstopdf}
\usepackage{epsfig}
\usepackage{color}
\usepackage{hyperref}
\def\be{\begin{equation}}
\def\ee{\end{equation}}
\def\bea{\begin{eqnarray}}
\def\eea{\end{eqnarray}}

\newcommand{\bigO}{\mathcal{O}}
\newcommand{\compC}{\mathbb{C}}
\let\Im\relax
\let\Re\relax
\DeclareMathOperator{\Im}{Im}
\DeclareMathOperator{\Re}{Re}
\DeclareMathOperator{\eff}{eff}

\begin{document}
\title{Generalized random matrix model with additional interactions}
\author{Swapnil Yadav, Kazi Alam and K. A. Muttalib}
\address
{Department of Physics, University of Florida, Gainesville, FL 32611-8440, USA}
\ead{\mailto{yadavswap.12@ufl.edu}, \mailto{kazi.a.alam@ufl.edu} and \mailto{muttalib@phys.ufl.edu}}
\author{Dong Wang}
\address
{Department of Mathematics, National University of Singapore, Singapore}
\ead{\mailto{matwd@nus.edu.sg}}

\begin{abstract}
We introduce a log-gas model that is a generalization of a random matrix ensemble with an additional interaction, whose strength depends on a parameter $\gamma$. The equilibrium density is computed by numerically solving the Riemann-Hilbert problem associated with the ensemble. The effect of the additional parameter $\gamma$ associated with the two-body interaction  can be understood in terms of an effective $\gamma$-dependent single-particle confining potential.

\end{abstract} 
\today

\maketitle

\section{Introduction}
 
The random matrix ensembles (see e.~g.~\cite{Mehta04,Erdos-Yau17}) introduced to explain the nuclear energy-level fluctuations are characterized by the joint probability density function (jpd) of the eigenvalues
\begin{equation}
  p(\{x_i\}) \propto \prod_{i=1}^Nw(x_i)\prod_{i<j}|x_i-x_j|^{\beta},\quad
  w(x)=e^{-V(x)} \text{ or } w(x) = e^{-NV(x)}, \quad \beta = 1, 2, 4,
\label{WD}
\end{equation}
where $\beta=2$ for unitary ensembles. Throughout this paper, we assume the convention $w(x) = e^{-NV(x)}$, so that the empirical distribution of the particles (aka equilibrium measure) converges as $N \to \infty$.
It is useful to describe the jpd in terms of an effective `Hamiltonian' $H$ of the eigenvalues defined by $p=\exp(-\beta H)$, where the term $\ln|x_i-x_j|$ in $H$ corresponds to a ``two-body interaction'' of a log-gas system, while the term $\frac{1}{\beta}V(x)$ corresponds to a single particle ``confining potential'' (see e.~g.~\cite{Forrester10}).  

As a toy model for quasi one-dimensional (1D) disordered conductors \cite{Beenakker97}, a solvable random matrix model with an additional two-body interaction was proposed in \cite{Muttalib95}, 
\begin{equation}
  p(\{x_i\};\theta) \propto \prod_{i=1}^Nw(x_i)\prod_{i<j}|x_i-x_j||x_i^{\theta}-x_j^{\theta}|, \quad 0 < \theta < \infty.
 \label{MB}
\end{equation}
This model was studied in detail by Borodin \cite{Borodin99}, and has become known as the Muttalib-Borodin (MB) ensemble \cite{Forrester-Wang15,Zhang15,Kuijlaars-Molag19}.
The special case of $\theta=2$ was later considered in \cite{Lueck-Sommers-Zirnbauer06} as a model of disordered bosons.  

It has later been argued that in contrast to a quasi 1D system, describing a three-dimensional (3D) disordered conductor  with appropriate eigenvector correlations needs a disorder-dependent parameter $\gamma$ that controls the strength of the two-body interaction \cite{Klauder-Muttalib99,Gopar-Muttalib02,Douglas-Markos-Muttalib14,Markos-Muttalib-Wolfle05}. 
The generic form that captures the essential features of this quasi 1D to 3D generalization has been suggested to be of the form
\begin{equation}
p(\{x_i\};\gamma) \propto \prod_{i=1}^Nw(x_i)\prod_{i<j}|x_i-x_j||r(x_i)-r(x_j)|^{\gamma}, \quad 0< \gamma \le 1,
\label{model-g}
\end{equation}
where $r(x)$ and $w(x)$ are appropriate functions relevant for disordered conductors \cite{Markos-Muttalib-Wolfle05}. As a solvable toy model that allows us to explore and study the role of the parameter $\gamma$,  we propose to investigate the simplest generalization of the MB ensemble, with $r(x)=x^{\theta}$ and $V(x)=2x$:
\begin{equation}
p(\{x_i\};\theta,\gamma) \propto \prod_{i=1}^Nw(x_i)\prod_{i<j}|x_i-x_j||x_i^{\theta}-x_j^{\theta}|^{\gamma},\quad 0< \gamma \le 1. 
\label{model}
\end{equation}
In particular, we will consider the case $\theta=2$ in detail, although the method is applicable for any $\theta>1$ and for any well behaved external confining potential. We will be interested in the case $x_i\ge 0$, since the transmission eigenvalues are non-negative \cite{Muttalib-Pichard-Stone87}.
We will call it the $\gamma$-ensemble. Note that $\gamma=1$ is just the MB ensemble of Eq.~(\ref{MB}). 

By solving the Riemann-Hilbert (RH) problem \cite{Deift99} that is  associated with certain integral transforms, see (\ref{complex_transforms}), of the (limiting) density of eigenvalues, Claeys and Romano, henceforth referred to as CR \cite{Claeys-Romano14}, have obtained the density of eigenvalues for the MB ensembles (Eq.~(\ref{MB})) for a linear as well as a quadratic potential, which have power-law divergences at the hard edge for all $\theta > 1$. In this work we generalize the method developed by CR  to the case of the $\gamma$-ensemble (Eq.~(\ref{model})) and study the density as a function of $\gamma$. Our results suggest that the $\gamma$-ensemble can be mapped on to an MB ensemble by replacing the single particle confining potential $V(x)$ with a $\gamma$-dependent effective potential $V_{\eff}(x;\gamma)$. This allows us to calculate the density for arbitrary values of $\gamma$. In particular we will show that as $\gamma$ is systematically reduced from $1$, the exponent of the diverging density at the hard edge changes from $-1/3$ for $\gamma=1$ (the MB ensemble) to $-1/2$ for $\gamma=0$ (the orthogonal Laguerre ensemble).

For the sake of completeness, we will repeat the method to study the effect of $\gamma$ on a model with non-diverging density, that is, with no hard edge. In particular we will apply the method to consider a model with a different two-body interaction, $r(x)=e^x$  with $-\infty < x < +\infty$, where the corresponding density has two soft edges. This shows that as long as the Joukowsky Transformation (JT) is known, the method can be applied to a wide variety of generalized models.

The paper is organized as follows. In Section \ref{sec:2} we briefly outline the equilibrium problem and the JT following CR. In Sections \ref{sec:3} and \ref{sec:4} we show how the method of CR can be adapted for the $\gamma$-ensembles to obtain the effective potential and the level density.  In Section \ref{sec:5} we use $V(x)=2x$ to show how the effective potentials and the corresponding level-densities change as $\gamma$ is reduced from 1 towards zero. Finally in Section \ref{sec:6}  we show briefly how the method can be applied to the case of $r(x)=e^x$ and $V(x)=\frac{x^2}{2}$ for which  the JT was obtained by Claeys and Wang \cite{Claeys-Wang11}, henceforth referred to as CW, and the density is non-diverging. Details of this model are provided in the \hyperref[Appendix]{Appendix}.


\section{The equilibrium problem for $\gamma = 1$} \label{sec:2}

This section is based on \cite{Claeys-Romano14}, and we borrow notation from there.

In terms of the Hamiltonian in (\ref{model}), by potential theory, particularly by an argument similar to that in \cite[Section 6.2]{Deift99}, there exists a unique equilibrium measure $\mu$ that minimizes the energy functional
\begin{equation}
  \frac{1}{2} \iint \ln \frac{1}{\lvert x - y \rvert} d\mu(x) d\mu(y) + \frac{\gamma}{2} \iint \ln \frac{1}{\lvert x^{\theta} - y^{\theta} \rvert} d\mu(x) d\mu(y) + \int V(x) d\mu(x),
\end{equation}
which satisfies the Euler-Lagrange (EL) equation
\begin{equation}
\int \ln| x - y| d\mu(y) +\gamma\int \ln| x^{\theta}-y^{\theta}|d\mu(y) 
- V(x)= \ell
\label{euler-lagrange}
\end{equation}
if $x$ lies inside the support of density. Here $\ell$ is some constant. Also the empirical distribution of the particles with jpd (\ref{model}) converges to this equilibrium measure. The equality sign is replaced by  $<$ if $x$ lies outside the support. The equilibrium problem for $\gamma=1$ has been solved exactly in CR under the ``one-cut'' condition that requires the equilibrium measure to be supported on a single interval, in the form of $[0, b]$ with $b > 0$ (``hard edge'' case) or $[a, b]$ with $0 < a < b$ (``soft edge'' case). In that case, if we define
\begin{equation}
\begin{aligned}
  g(z) \equiv {}& \int \log(z-x)d\mu(x), && z \in \mathbb{C}\backslash(-\infty,b], \\
  \tilde{g}(z) \equiv {}& \int \log(z^{\theta}-x^{\theta})d\mu(x), && z \in \mathbb{H}_\theta\backslash(0,b],
\end{aligned}
\label{complex_transforms}
\end{equation}
where $\mathbb{H}_{\theta}$ is defined as
\begin{equation}
  \mathbb{H}_{\theta} = \{ z \in \compC \mid -\frac{\pi}{\theta} < \arg(z) < \frac{\pi}{\theta} \},
\end{equation}
then the equilibrium measure $\mu$ can be characterized by a vector-valued Riemann-Hilbert (RH) problem. Since this paper concentrates on the hard edge case, we only state the RH problem in the hard edge case:

\paragraph{RH problem for $(g, \tilde{g})$} ($\gamma = 1$)

\begin{itemize}
\item
  $(g, \tilde{g})$ is analytic in $(\compC \setminus (-\infty, b], \mathbb{H}_{\theta} \setminus [0, b])$.
\item
  Writing $g_+$, $g_-$, $\tilde{g}_+$, $\tilde{g}_-$ for the boundary values of $g$ and $\tilde{g}$ when approaching $(-\infty, b)$ and $(0, b)$ from above (for $+$) or below (for $-$), we have the relations
  \begin{subequations}
    \begin{align}
      g_{\pm}(x) + \tilde{g}_{\mp}(x) = {}& V(x) + \ell, && \text{for $x \in (0, b)$}, \label{eq:RHPa} \\
      \tilde{g}(e^{-i\pi/\theta} x) = {}& \tilde{g}(e^{i\pi/\theta} x) - 2\pi i, && \text{for $x > 0$}, \label{eq:RHPb} \\
      g_+(x) = {}& g_-(x) + 2\pi i, && \text{for $x < 0$}. \label{eq:RHPc}
    \end{align}
  \end{subequations}
\item
  As $z \to \infty$ in $\compC$, $g(z) = \log(z) + \bigO(z^{-1})$ and as $z \to \infty$ in $\mathbb{H}_{\theta}$, $\tilde{g}(z) = \theta\log z + \bigO(z^{-\theta})$.
\end{itemize}
We can find the density function for $\mu$ by solving the RH problem above. In doing so, a crucial role is played by the Joukowsky Transformation (JT) for the hard edge case
\begin{equation}
\begin{aligned}
  J_c(s) = {}& c(s+1)(\frac{s+1}{s})^{\frac{1}{\theta}}, \\
\end{aligned}
\label{joukowsky}
\end{equation}
where $s$ is a complex variable, and the parameter $c$ depends on $b$ such that $b=c\frac{(1+\theta)^{1+\frac{1}{\theta}}}{\theta}$. 

While the vector-valued RH problem and the JT in (\ref{joukowsky}) was obtained for $\gamma=1$, it turns out that the equilibrium problem for $\gamma <1$  can also be solved through them. In the following two sections we will briefly outline how the above RH problem and JT can be used to obtain the density function for arbitrary $0<\gamma <1$.


\section{Effective potential} \label{sec:3}

To accommodate $0<\gamma <1$ for non-negative eigenvalues within the CR framework, we consider the hard edge case, focusing on $\theta=2$ for simplicity.

\begin{figure}
\begin{center}
\includegraphics[width=0.48\textwidth]{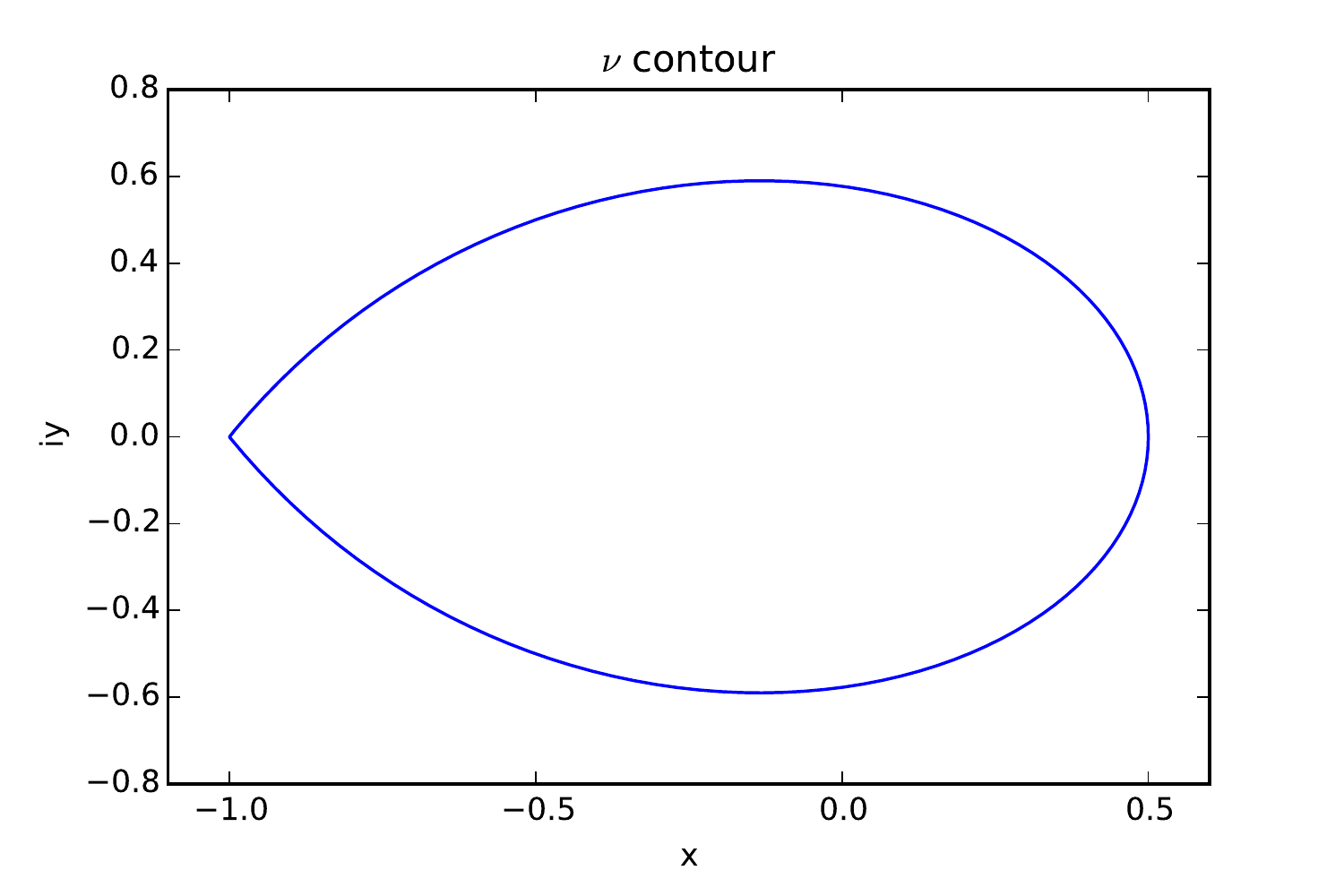}
\end{center}
\caption{
(Color online) 
$\nu$ contour for $\theta=2, \ c=1 $.}
\label{contour_CR}
\end{figure}

\begin{figure}
\begin{center}
\includegraphics[width=0.48\textwidth]{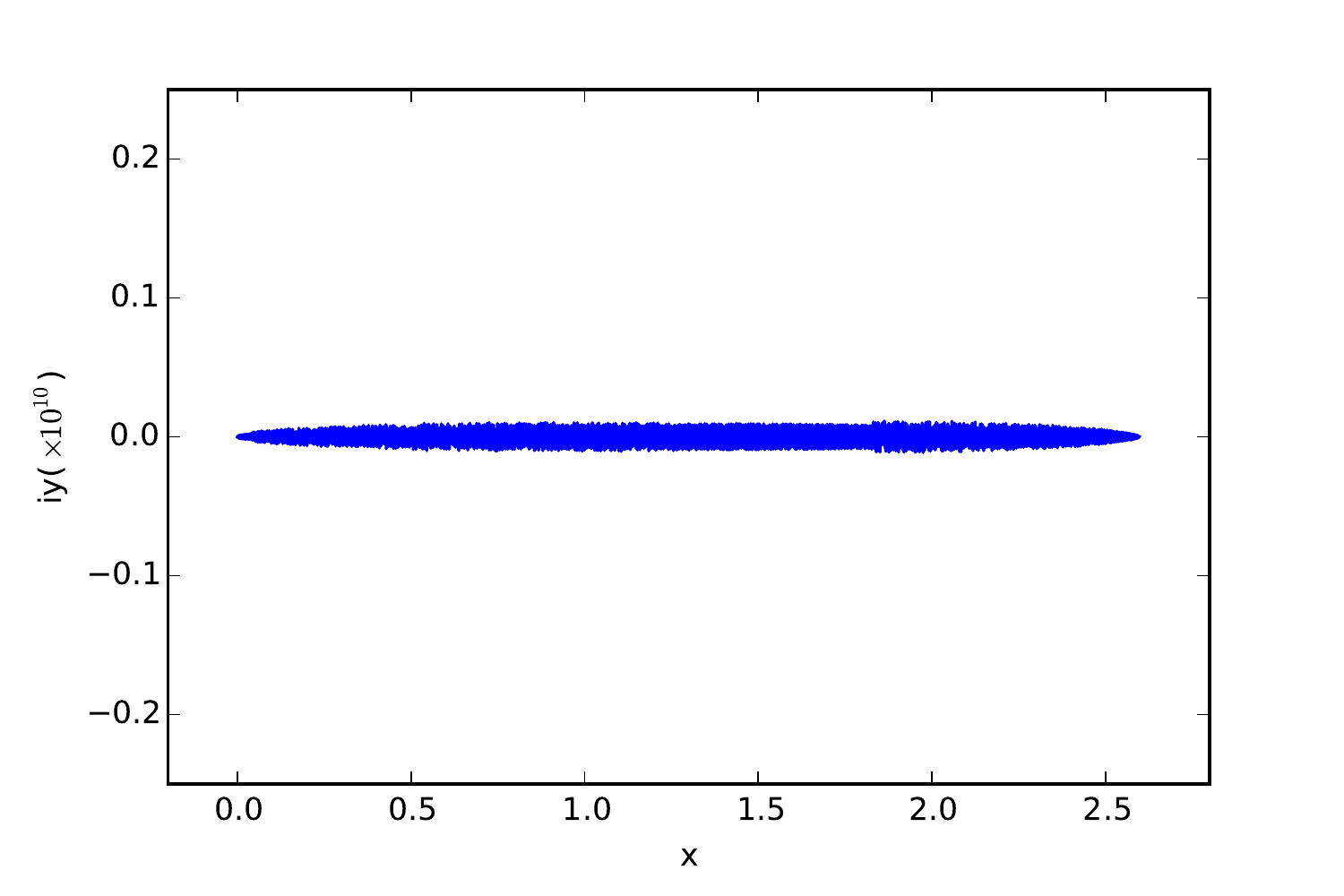}
\end{center}
\caption{
(Color online) 
Mapping for $\nu_1$ contour, $\theta=2, \ c=1 $. Mapping for $\nu_2$ looks similar. }
\label{mapping_CR}
\end{figure}

\begin{figure}
\begin{center}
\includegraphics[width=0.48\textwidth]{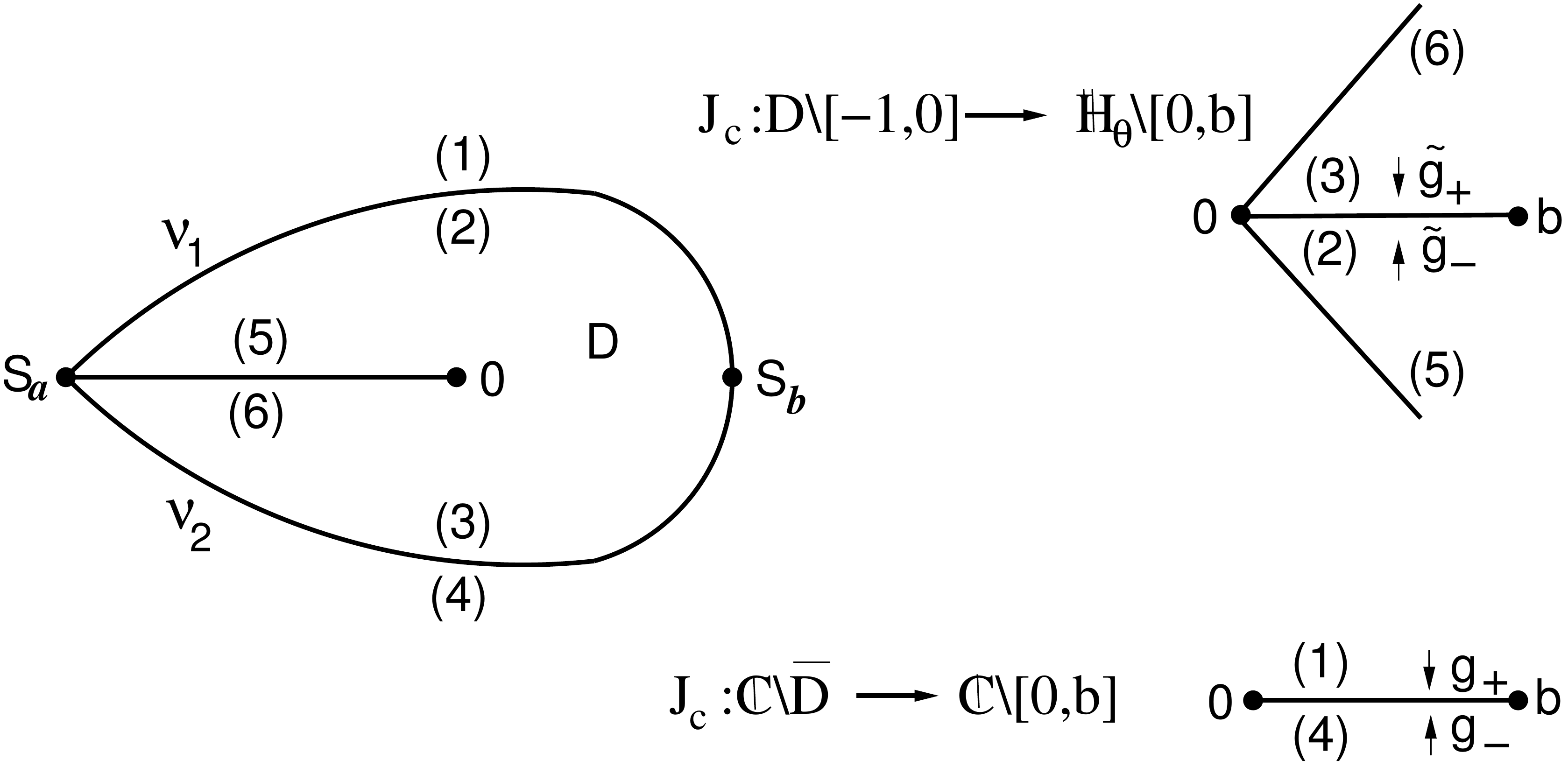}
\end{center}
\caption{
(Color online) 
Schematic Figure for mapping of JT, following CR.}
\label{mapping_schematic}
\end{figure}

A closer look of the JT for hard edge (\ref{joukowsky}) shows that it is analytic in $\mathbb{C}\backslash[-1,0]$ and has critical points on real line at $S_a=-1$ and $S_b=\frac{1}{\theta}$ which are mapped to points $0$ and $b$, respectively. There also exist points in the complex plane which are mapped on to the real line between $0$ and $b$ by $J_c(s)$. The equation of locus of such points is given by 
\begin{equation}
r(\phi)= \left. \tan \left( \frac{\phi}{1+\theta} \right) \middle/ \left[ \sin\phi-\cos\phi\tan \left( \frac{\phi}{1+\theta} \right) \right] \right.,
\end{equation} 
where $0<\phi<2\pi$ is the argument of point $s$ in the complex plane. This defines a closed contour $\nu$ in the complex plane which is symmetric about the $x$-axis. We denote the two symmetric parts as curves $\nu_1$ (upper) and $\nu_2$ (lower) which are complex conjugates of each other.
Figure \ref{contour_CR} and Figure \ref{mapping_CR} show contour $\nu$ for $\theta=2,c=1$ and its mapping, respectively. Since this mapping calculation is numerical, in Figure \ref{mapping_CR} we see very small $y$ components as well. In this paper we orient $\nu$ positively, so $\nu_1$ is from right to left and $\nu_2$ is from left to right.
In Figure \ref{mapping_schematic} we show details of this mapping  schematically. In particular,  all points except the branch cut $[-1,0]$ in the region $D$ inside the contour $\nu$ is mapped on to a complex region $\mathbb{H}_\theta\backslash[0,b]$, while all outside points are mapped on to a different complex region $\mathbb{C}\backslash [0,b]$.

To solve for $\mu(x)$ using the EL equations we define complex transforms $g$ and $\tilde{g}$ as in \eqref{complex_transforms}. Here $(g,\tilde{g})$ is analytic in $(\mathbb{C}\backslash(-\infty,b],\mathbb{H}_\theta\backslash(0,b])$ respectively so that the logarithms are well defined. Let $g_+,g_-$ and $\tilde{g}_+,\tilde{g}_-$ denote boundary values of $g$ and $\tilde{g}$ when approaching $(-\infty,b]$ in $\mathbb{C}$  and $(0,b]$ in $\mathbb{H}_\theta$ respectively from above $(+)$ and below $(-)$, the same as the notation used in Section \ref{sec:2}, also as shown schematically in Figure \ref{mapping_schematic}.

Our $g$ and $\tilde{g}$ satisfy the RH problem stated in Section \ref{sec:2} in the $\gamma = 1$ case, except for property \eqref{eq:RHPa}, which is generalized to
\begin{equation} \label{eq:generalized_jump}
  g_{\pm}(x)+\gamma\tilde{g}_{\mp}(x)-V(x)-\ell = \pm\pi i (1 - \gamma) \mu([x, b]),
\end{equation}
because for $x \in [0, b]$
\begin{equation}
  g_{\pm}(x) = \int^b_0 \ln \lvert x - y \rvert d\mu(y) \pm \pi i \mu([x, b]), \quad \tilde{g}_{\pm}(x) = \int^b_0 \ln \lvert x^{\theta} - y^{\theta} \rvert d\mu(y) \pm \pi i \mu([x, b]).
\end{equation}
Rewriting $g=(1+\gamma)g/2 +(1-\gamma)g/2$, we have that \eqref{eq:generalized_jump} is equivalent to
\begin{equation}
\bigg{(}\frac{1\pm\gamma}{2}\bigg{)}g_+(x)+\bigg{(}\frac{1\mp\gamma}{2}\bigg{)}g_-(x)+\gamma\tilde{g}_{\mp}(x) 
=V(x)+\ell.
\label{euler-lagrange_g} 
\end{equation}
Following CR, we define $G(s)\equiv g^{\prime}(s)$ and $\tilde{G}(s)\equiv \tilde{g}^{\prime}(s)$ where the prime denotes derivative with respect to its argument.
Also define
\begin{equation}
M(s)\equiv
\begin{cases}
  G(J_c(s)), & \text{for } s\in\mathbb{C}\backslash \bar{D}, \\
  \tilde{G}(J_c(s)), & \text{for } s\in D\backslash[-1,0],
\end{cases}
\label{M_def}    
\end{equation}
where $D$ is the domain inside $\nu$, as shown in Figure \ref{mapping_schematic}.
For $x \in (0, b)$, there are $s_1 \in \nu_1$ and $s_2 \in \nu_2$ such that $J_c(s_1) = J_c(s_2) = x$. Then for $g_+(x)$ in Eq.~(\ref{euler-lagrange_g}), it is equal to the limit of $g(J_c(s))$ as $s\rightarrow s_1 \in \nu_1$ from outside of contour $\nu$ (see Figure \ref{mapping_schematic}). Similarly for $\tilde{g}_+(x)$, it is equal to the limit of $\tilde{g}(J_c(s))$ as $s\rightarrow s_2 \in \nu_2$ from inside of contour $\nu$. Hence by taking derivative, the properties of $g_{\pm}(x)$ above implies the properties of $M(s)$ 
\begin{equation}
\begin{split}
  M_+(s_1)+\gamma M_-(s_1)+M_-(s_2)+\gamma M_+(s_2) 
  = {}& 2V^{\prime}(J_c(s)), \\
  M_+(s_1)-M_-(s_2)+M_-(s_1)-M_+(s_2) 
  = {}& 0.
\end{split}
\label{sum-difference}
\end{equation}
Following CR we define 
$
N(s)\equiv M(s)J_c(s)
$,
so (\ref{sum-difference}) can be rewritten in terms of $N(s)$ and $J_c(s)$. In addition, $J_c(s^+_1) = J_c(s^-_1) = x$ where $J_c(s^+_1)$ (resp.\ $J_c(s^-_1)$) is the limit of $J_c(s)$ with $s$ approaching $s_1 \in \nu_1$ from outside (resp.\ inside) of $\nu$(see Figure \ref{mapping_schematic}). Thus we can replace both $J_c(s^+_1)$ and $J_c(s^-_1)$ by $J_c(s)=x$. We have

\paragraph{RH problem for $N$}

\begin{itemize}
\item
  $N$ is analytic in $\compC \setminus \nu$.
\item 
 \begin{equation}
    \begin{split}
      N_+(s_1)+\gamma N_-(s_1)+N_-(s_2)+\gamma N_+(s_2) 
      = {}& 2V^{\prime}(J_c(s))J_c(s), \cr  
      N_+(s_1)-N_-(s_2)+N_-(s_1)-N_+(s_2) 
      = {}& 0.
    \end{split}
    \label{Npm}
  \end{equation}
\item 
  $N(0) = \theta$ and $N(s) \to 1$ as $s \to \infty$.
\end{itemize}
Suppose we have a function $f$ such that for all $s \in \nu$
\begin{equation} \label{eq:defn_f}
f(J_c(s))\equiv N_+(s)+N_-(s).
\end{equation}
From the RH problem above that $N$ satisfies, we find the solution to $N(s)$ as
\begin{equation}
N(s)=
\begin{cases}
  \frac{-1}{2\pi i}\oint_{\nu}\frac{f(J_c(\xi))}{\xi -s}\; d\xi +1, & s\in \mathbb{C}\backslash \bar{D}, \cr
\frac{1}{2\pi i}\oint_{\nu}\frac{f(J_c(\xi))}{\xi -s}\; d\xi -1, & s\in D\backslash [-1,0]
\end{cases}
\label{N_def_contr}
\end{equation}
where contour $\nu$ is for JT $J_c(s)$ \cite{Claeys-Romano14}. Also from the RH problem, we find that the constant $c$ in this JT satisfies the equation
\begin{equation}
\label{c_hard_edge}
\frac{1}{2\pi i}{\displaystyle \oint_{\nu}^{}}\frac{f(J_c(s))}{s}ds=1+\theta.
\end{equation}

It is clear now that if we have a well defined function $f$ that satisfies \eqref{eq:defn_f}, then we can find $N(s)$, or equivalently $g(z)$ and $\tilde{g}(z)$ explicitly, and finally have a formula for the dentity function of equilibrium measure $\mu$. Below we explain the main technical contribution of this paper, the numerical method to find $f$.

Equation (\ref{Npm}) can now be rewritten as
\begin{equation}
(1-\gamma)(N_+(s_1)+N_-(s_2))+2\gamma f(J_c(s)) 
=2V^{\prime}(J_c(s))J_c(s). 
\end{equation}
From Equation(\ref{N_def_contr}) we have,
\begin{equation}
N_+(s_1)=\frac{1}{2 \pi i}\oint_\nu \frac{f(J_c(s))}{(s_1)_+ - s}ds +1, \quad
N_-(s_2)=\frac{1}{2 \pi i}\oint_\nu \frac{f(J_c(s))}{(s_2)_- - s}ds +1 .
\label{N_pm_def_contr}  
\end{equation}

Let us now define the inverse mapping of $J_c$ as
\begin{equation}
s=J_c^{-1}(x)=h(x).
\label{inversemap}
\end{equation}
It is generally double-valued, and we can take the appropriate one.
Note that for both $N_+(s_1)$ and $N_-(s_2)$ in Eq.~(\ref{N_pm_def_contr}), the function is defined by the limit of $N(s)$ as $s$ approaches $s_1$ or $s_2$ on $\nu$ from outside. Hence we used the first identity in Eq.~(\ref{N_def_contr}). Let $(s_1)_+ = h(y) \ ; \ (s_2)_- = \bar{h}(y) \ ; \ s_1 = h(x) \ \text{and} \ s_2 = \bar{h}(x) $ where the bar denotes complex conjugate. ($h(y) - h(x)$ is infinitesimal if $y = x$, but it is crucial that $h(y)$ is outside of $\nu$ while $h(x)$ is on $\nu$.) Writing Eq.~(\ref{N_pm_def_contr}) in terms of the inverse mappings we get
\begin{equation}
\begin{split}
  N_+(s_1)= {}& \frac{1}{2 \pi i}\int_{\nu_1} \frac{f(x)}{h(y) - h(x)}dh(x)  
  + \frac{1}{2 \pi i}\int_{\nu_2} \frac{f(x)}{h(y) - \overline{h}(x)}d\overline{h}(x) +1, \\
  N_-(s_2)= {}& \frac{1}{2 \pi i}\int_{\nu_1} \frac{f(x)}{\overline{h}(y) - h(x)}dh(x)  
  + \frac{1}{2 \pi i}\int_{\nu_2} \frac{f(x)}{\overline{h}(y) - \overline{h}(x)}d\overline{h}(x) +1.                
\end{split}
\end{equation}

Recall that $\nu_1$ is oriented from $S_b$ to $S_a$. Thus in the  mapped space, limits of the corresponding real integral are from $b$ to $0$. Similarly for $\nu_2$, the real integral is from $0$ to $b$. Combining the two, writing the integrals in the mapped real space and  
substituting for $[N_+(s_1)+N_-(s_2)]$ we finally get the integral equation for $f$,
\begin{equation}
\label{f_integral_eqn}
  f(y;\gamma) = \frac{V^{\prime}(y)y}{\gamma} 
  -\frac{1-\gamma}{\gamma}\bigg[1 + \frac{1}{2\pi}\int_0^b f(x;\gamma)\phi(x,y)dx \bigg],
\end{equation}
where
\begin{equation}
  \phi(x,y) = \Im\bigg[ \left( \frac{1}{h(y) - \overline{h}(x)} + \frac{1}{\overline{h}(y) - \overline{h}(x)} \right) \overline{h}^{\prime}(x) \bigg].
\end{equation}
We solve the above integral equation  (\ref{f_integral_eqn})  for $f(y; \gamma)$ and Eq.~(\ref{c_hard_edge}) for $c$  numerically self-consistently.

Using the definition for $f(x;\gamma)$ we further find the new effective potential $V_{\eff}(x;\gamma)$ which is related to $f(x;\gamma)$ by 
\begin{equation}
V'_{\eff}(x;\gamma)=\frac{f(x;\gamma)}{x}.
\label{V-effective}
\end{equation} 
This is one of the central results of this work. It shows that at the global density level the $\gamma$-ensembles can be mapped onto an MB ensemble with an appropriate effective single-particle potential. Thus methods developed for studying the MB ensemble can be adapted to study the $\gamma$-ensembles.


\section{Level density} \label{sec:4}

With given definition of $V_{\eff}$, the constant $c$ for JT satisfies equation similar to the one in CR except that $V$ is now replaced by $V_{\eff}$. 

\begin{equation}
\label{c_hard_edge2}
\frac{1}{2\pi i}{\displaystyle \oint_{\nu}^{}}\frac{U_{c}(s)}{s}ds=1+\theta, \quad U_{c}(s)=V^{\prime}_{\eff}(J_{c}(s); \gamma)J_{c}(s)=f(J_{c}(s); \gamma).
\end{equation}
Then the density corresponding to the $\gamma$-ensembles is computed using the relation  \cite{Claeys-Romano14} $\sigma(y)=-[N_+(s_1)-N_-(s_2)]/2\pi iy$.   
Substituting for $N_+(s_1)$ and $N_-(s_2)$ using Eq.~(\ref{N_def_contr}), the expression for density becomes,
\begin{equation}
\label{density_hard_edge}
\begin{split}
  \sigma(y;\gamma)= {}& \frac{-1}{2{\pi}^2 \gamma y}\int_{b}^0 xV'_{\eff}(x;\gamma)\chi(x,y) dx,\\
  \chi(x,y)= {}& \Re\bigg[ \bigg( \frac{1}{\overline{h}(y) - h(x)}-\frac{1}{h(y) - h(x)} \bigg)h^{\prime}(x)\bigg].
\end{split}
\end{equation}
The inverse mappings $h$ and $\overline{h}$ are from complex mapping $[0,b]$ to the contour $\nu$. 
Comparing with CR, it shows that the density for $\gamma<1$ has the same expression as that for $\gamma=1$, except that the potential $V(x)$ is replaced by the corresponding effective potential $V_{\eff}(x;\gamma)$.


\section{Results for $\theta=2$} \label{sec:5}

\begin{figure}
\begin{center}
\includegraphics[width=0.48\textwidth]{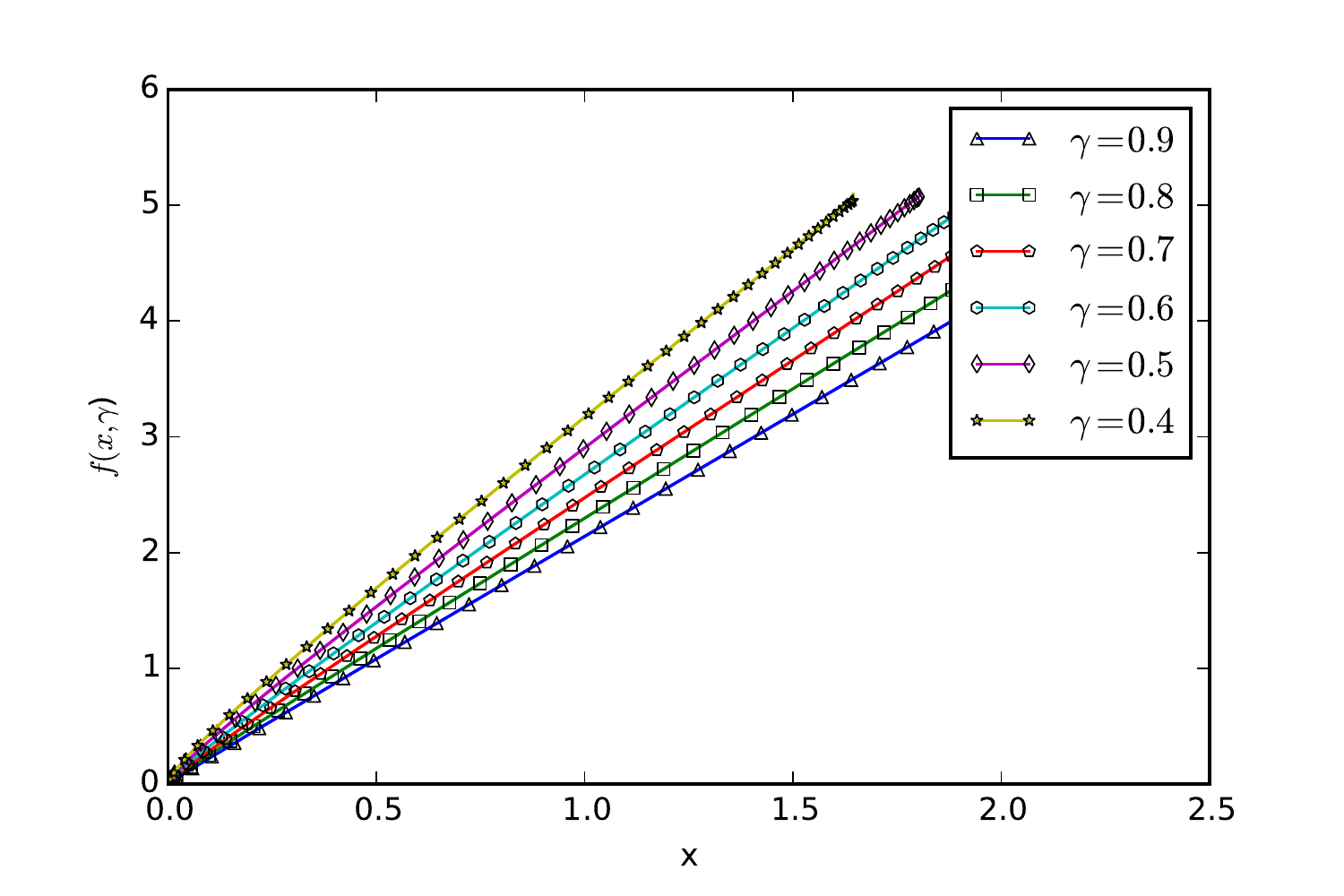}
\includegraphics[width=0.48\textwidth]{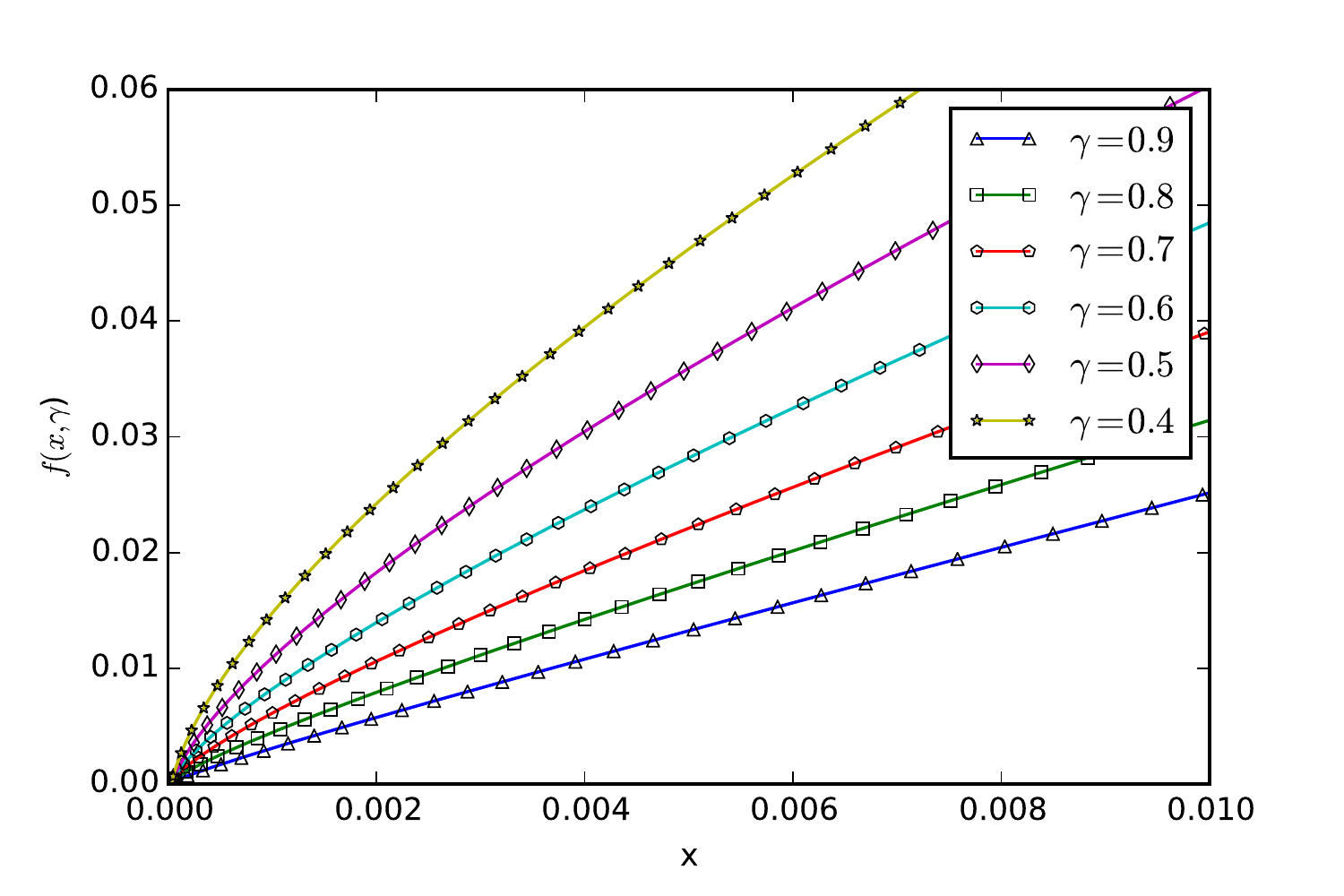}
\end{center}
\caption{
(Color online) 
Left panel: $f(x;\gamma)$ for different values of $\gamma$.
Right panel: Expanded view near origin.}
\label{f-x}
\end{figure}

The formulation developed so far is independent of the choice of the confining potential $V(x)$. As a concrete example, we consider a potential of the form
\begin{equation}
V(x)=tx.
\label{linear-potential}
\end{equation} 
We will choose $t=2$ as in CR. 
We consider the hard edge case for $\gamma < 1$ and $\theta=2$.  We solve the self-consistent integral equation (Eq.~(\ref{f_integral_eqn})) for $f(x;\gamma)$ numerically for different values of $\gamma$. Figure \ref{f-x} shows $f(x;\gamma)$ for selected values of $\gamma$.  Using the definition Eq.~(\ref{V-effective}), we computed the corresponding $V_{\eff}(x;\gamma)$ for each $\gamma$.
\begin{figure}
\begin{center}
\includegraphics[width=0.48\textwidth]{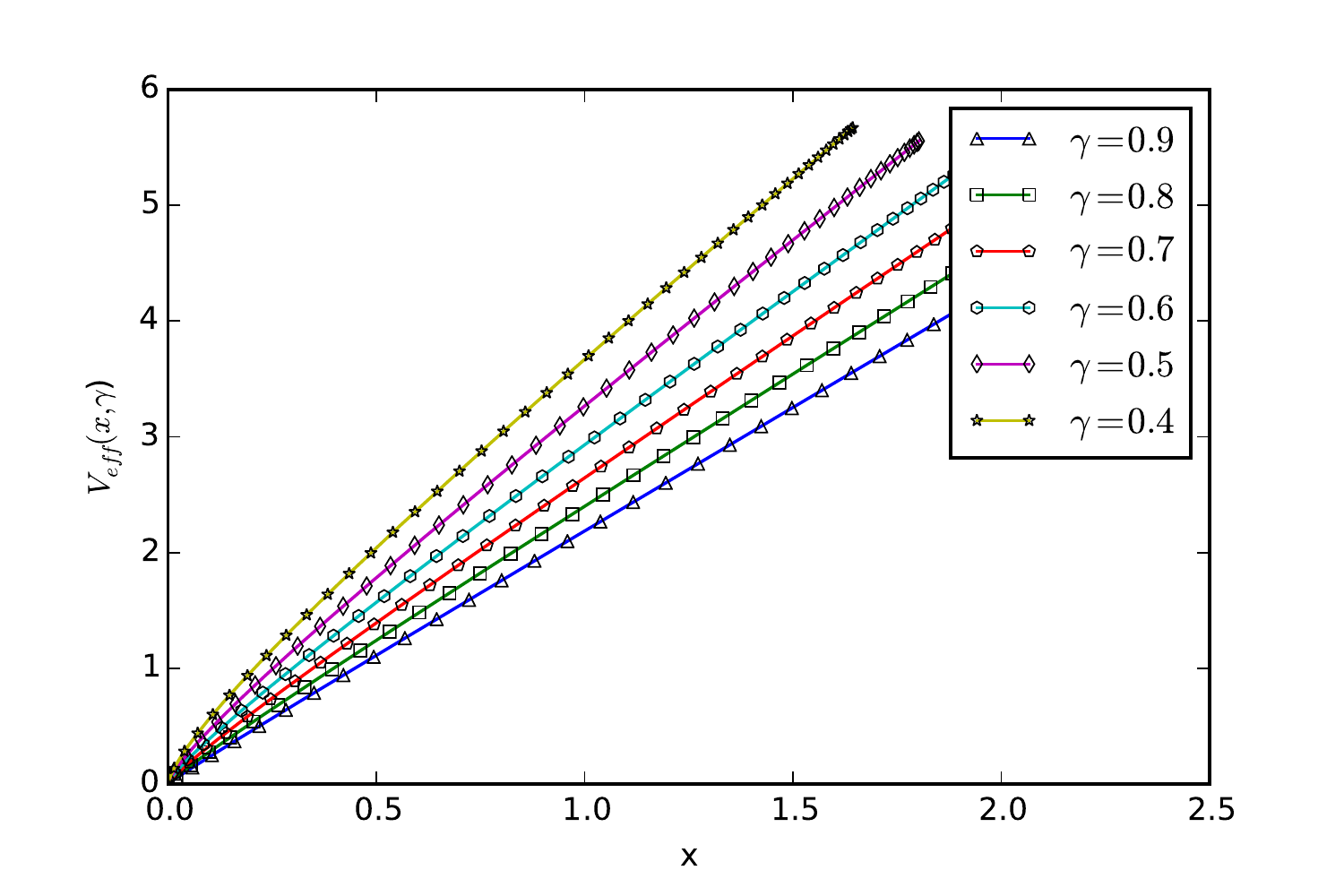}
\includegraphics[width=0.48\textwidth]{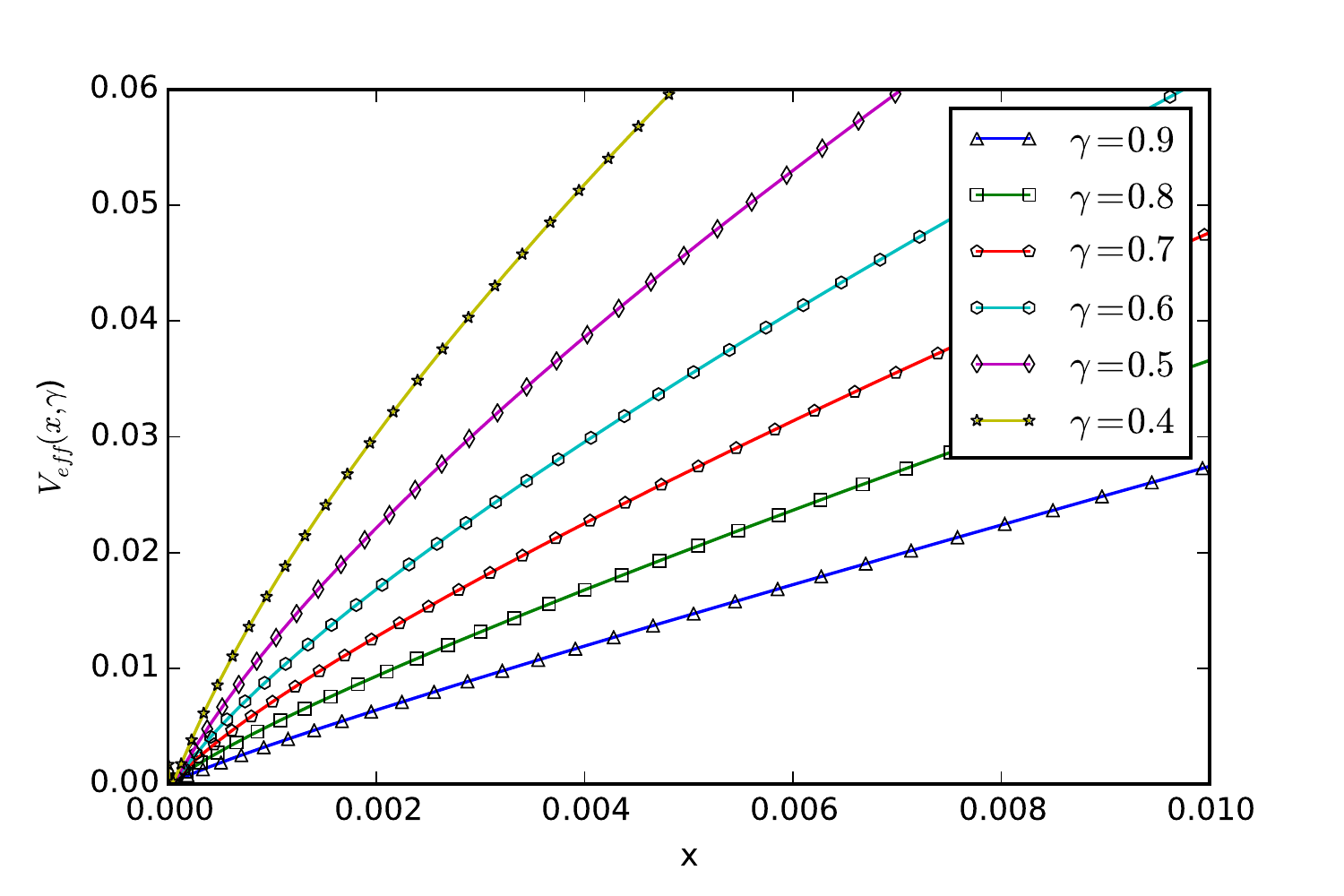}
\end{center}
\caption{
(Color online) 
Left panel: $V_{\eff}(x;\gamma)$ for different values of $\gamma$.
Right panel: Expanded view near origin.}
\label{V(x)-theta2-linear2}
\end{figure}
Figure \ref{V(x)-theta2-linear2} shows the results. 
\begin{figure}
\begin{center}
\includegraphics[width=0.48\textwidth]{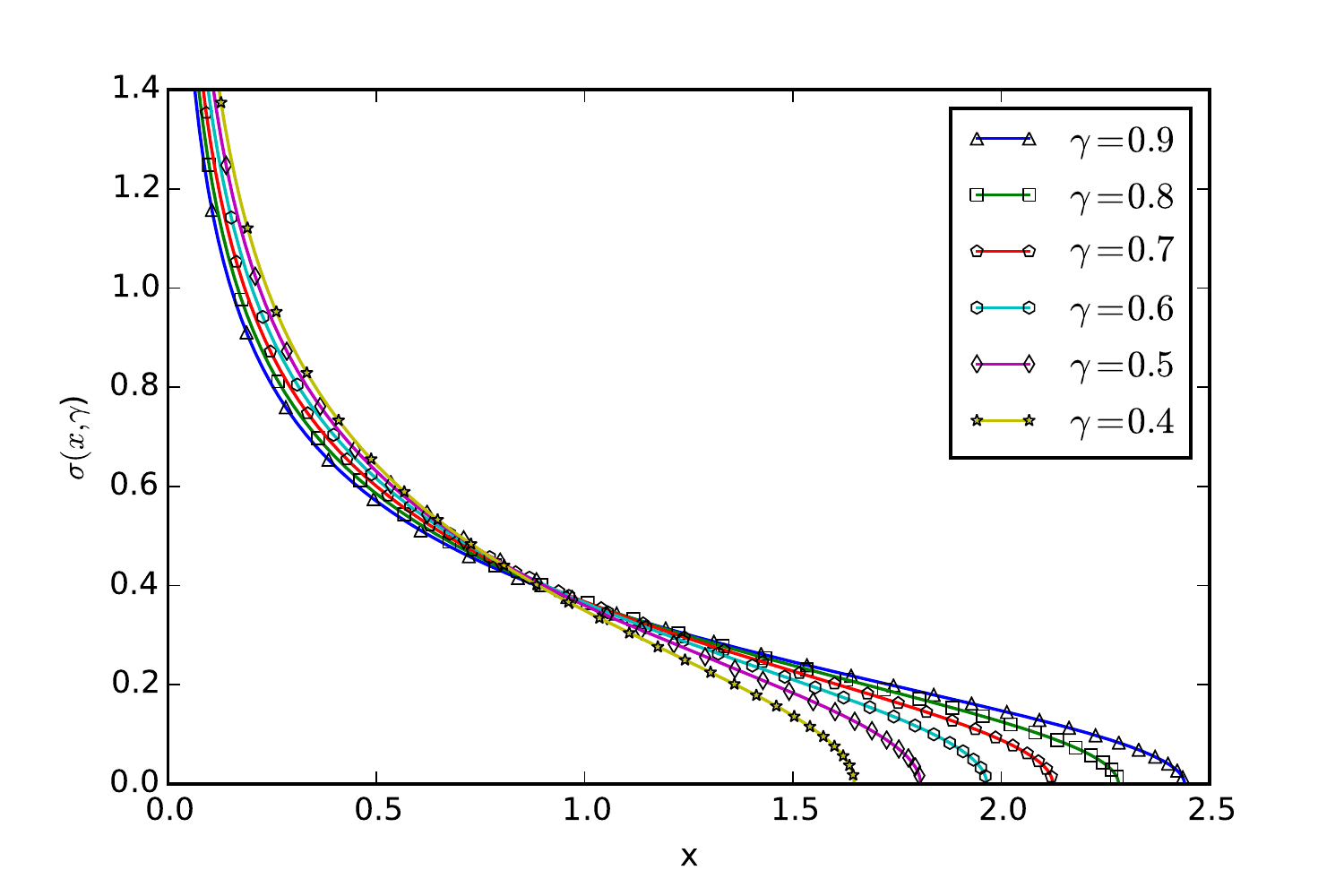}
\includegraphics[width=0.48\textwidth]{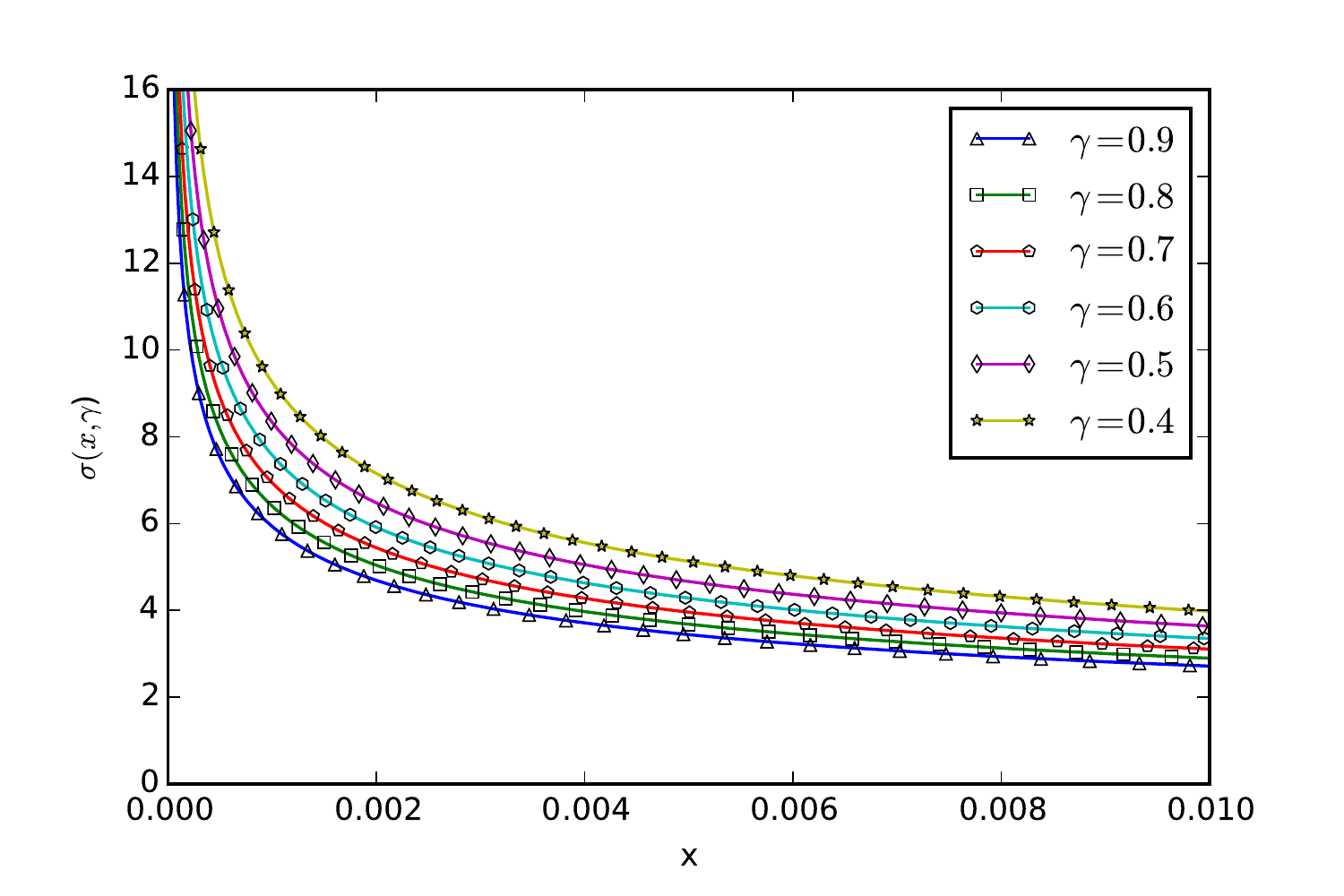}
\end{center}
\caption{
(Color online) Left panel:
Normalized density corresponding to $V_{\eff}$ in Figure \ref{V(x)-theta2-linear2}. Right panel: Expanded view near origin} 
\label{density-theta2-linear2}
\end{figure}

\begin{figure}
\begin{center}
\includegraphics[width=0.48\textwidth]{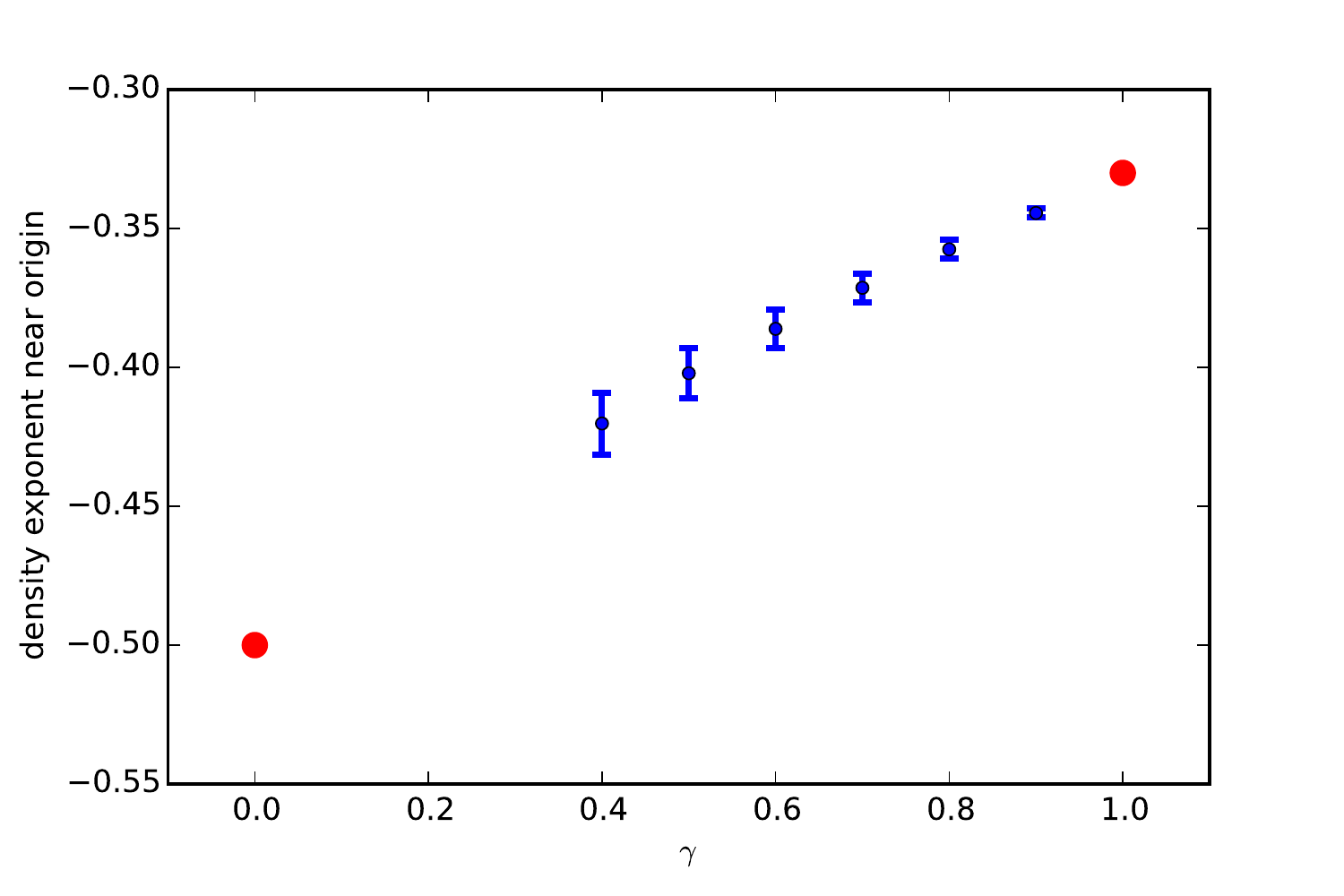}
\end{center}
\caption{
(Color online) Exponents with uncertainties in the numerical estimates. Points for $\gamma=1$ and $\gamma=0$ are known analytically.} 
\label{exponent}
\end{figure}

The densities evaluated from the effective potentials for different $\gamma$ are shown in Figure \ref{density-theta2-linear2}. The diverging exponent at the hard edge changes as a function of $\gamma$. Figure \ref{exponent} shows the crossover between the known exponents -1/3 for $\gamma=1$ and -1/2 for $\gamma=0$ as a function of $\gamma$.  


\section{Non-diverging density} \label{sec:6}

\begin{figure}
\begin{center}
\includegraphics[width=0.48\textwidth]{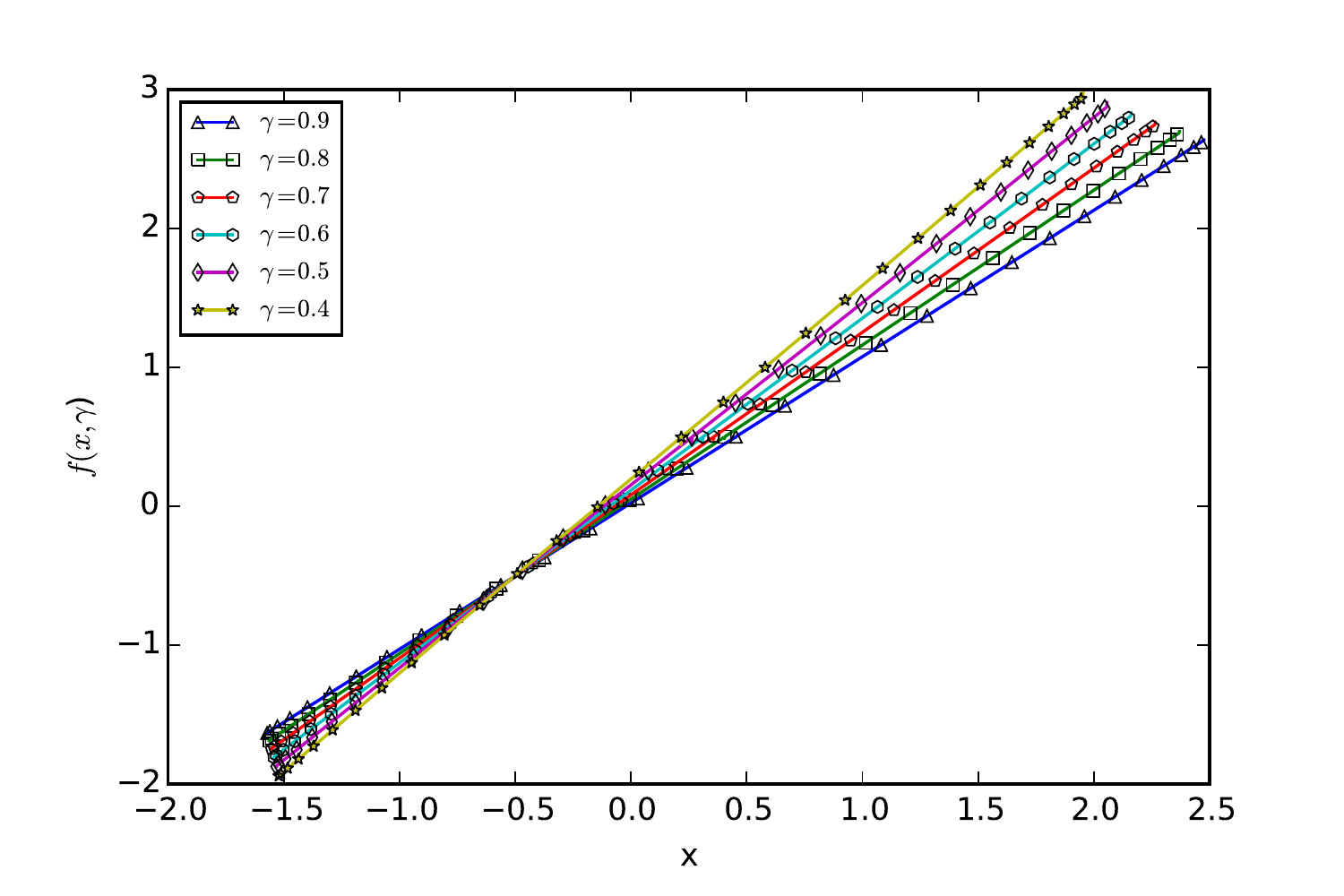}
\end{center}
\caption{
(Color online) 
$f_e(x;\gamma)$ for different values of $\gamma$.}
\label{f-x-e}
\end{figure}

\begin{figure}
\begin{center}
\includegraphics[width=0.48\textwidth]{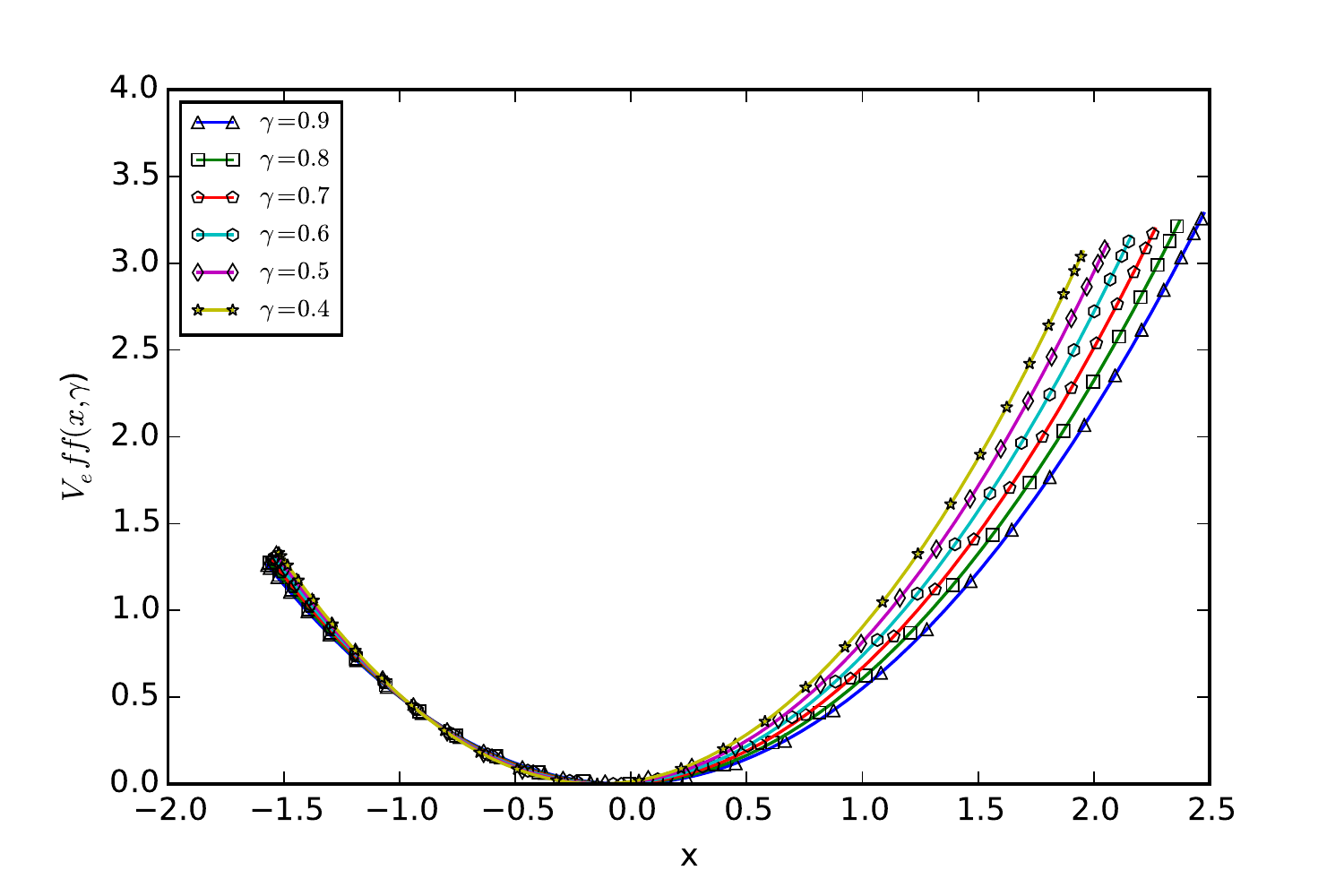}
\includegraphics[width=0.48\textwidth]{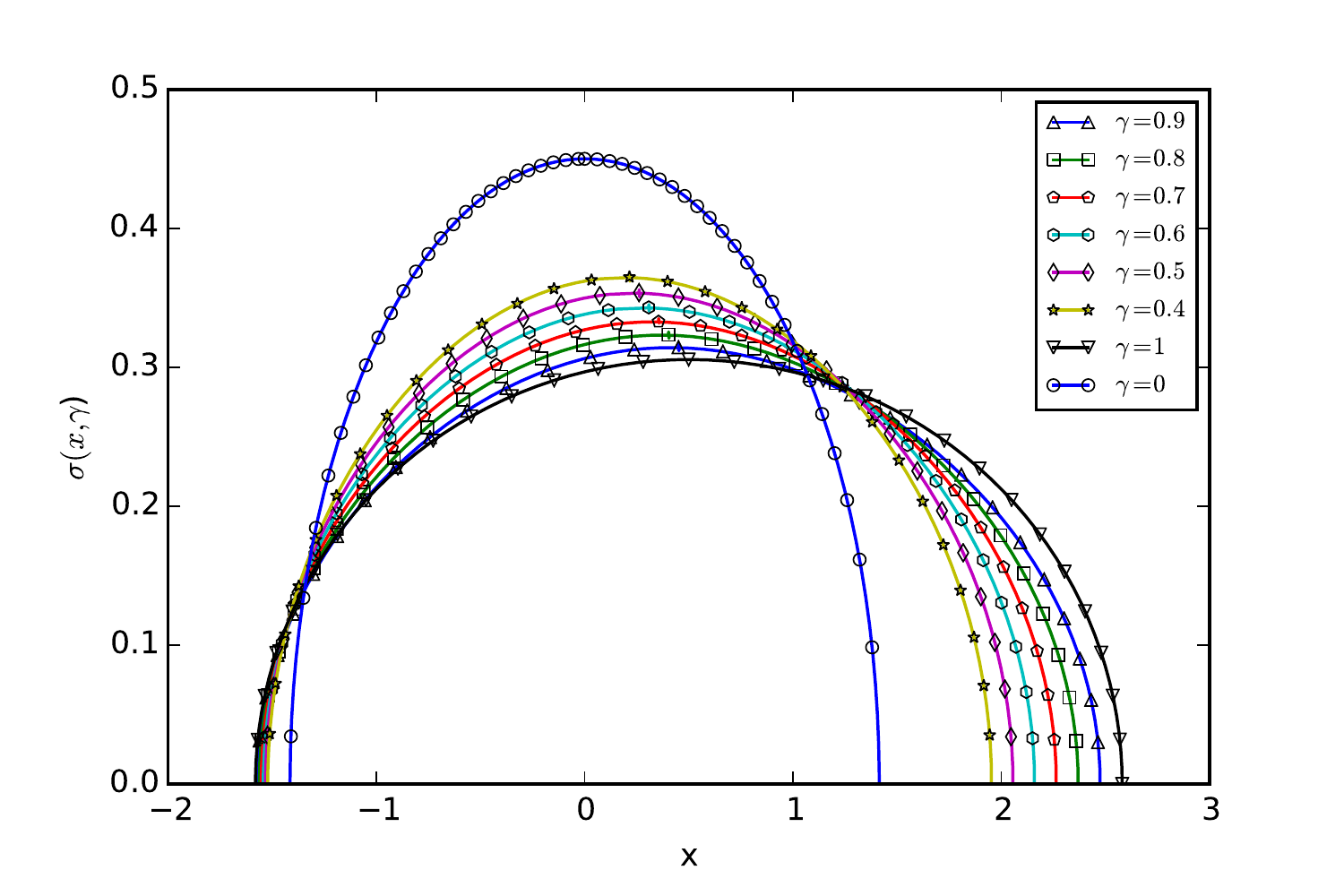}
\end{center}
\caption{
(Color online) 
The effective potential (Left panel) and the density (Right panel) for model (\ref{model-e}). Densities for $\gamma=1$ and $\gamma=0$ are known analytically.}.
\label{results-e}
\end{figure}

Finally, as an example of a model with non-diverging density which has two soft edges, we consider a $\gamma$-generalization of the model (\ref{model-g}) with $r(x)=e^x$ and $w(x)=e^{\frac{-Nx^2}{2}}$, where $-\infty < x < +\infty$:
\begin{equation}
p(\{x_i\}; \gamma) \propto \prod_{i=1}^Nw(x_i)\prod_{i<j}|x_i-x_j||e^{x_i}-e^{x_j}|^{\gamma},\quad 0< \gamma \le 1. 
\label{model-e}
\end{equation}
The model with $\gamma=1$ has been studied in detail by CW \cite{Claeys-Wang11}, who obtained the necessary JT. As with the generalized MB ensemble, we use the JT of CW and follow the method developed in Sections \ref{sec:3} and \ref{sec:4} to obtain the effective potential and hence the density for (\ref{model-e}) for different values of $\gamma$. We present the details in the \hyperref[Appendix]{Appendix}. The results for $f_e(x)$, the effective potentials and the densities for different values of $\gamma$ are given in Figures \ref{f-x-e} and  \ref{results-e}.
 
\section{Summary and conclusion} \label{sec:7}

We have introduced a toy model, Eq.~(\ref{model}), as a generalization of the MB random matrix  ensemble, Eq.~(\ref{MB}), with an additional parameter $\gamma$. This model is a solvable version of a realistic model for 3D conductors, albeit with a simplified two-body interaction. In order to solve for the density, we develop a method based on the solution of the associated RH problem, following CR. In principle, any two-body interaction can be solved provided the appropriate JT is known.  As an example, we also consider an interaction of the form $\ln |e^{x_i}-e^{x_j}|$ with $-\infty < x < +\infty$ for which the JT has been obtained by CW. It would be interesting to consider this latter model with a hard edge, in order to be able to compare how different two-body interactions affect the role of the parameter $\gamma$.

Our method exploits the fact that the effect of the parameter $\gamma$ can be understood in terms of an effective $\gamma$-dependent potential $V_{\eff}(x;\gamma)$, which replaces the starting confining potential $V(x)$. Hopefully, this will allow us to obtain not only the density, but also the two-level kernel from which correlations like the gap-function and the nearest-neighbor spacing distributions can be obtained. 

\section*{Acknowledgments}

KAM would like to thank the Department of Mathematics, NUS where he spent part of his sabbatical in 2017 and where this work originated. DW was partially supported by the Singapore AcRF Tier 1 grant R-146-000-217-112 (which partially supported KAM's visit) and the Chinese NSFC grant 11871425.


\section*{Appendix} \label{Appendix}

\begin{figure}
\begin{center}
\includegraphics[width=0.48\textwidth]{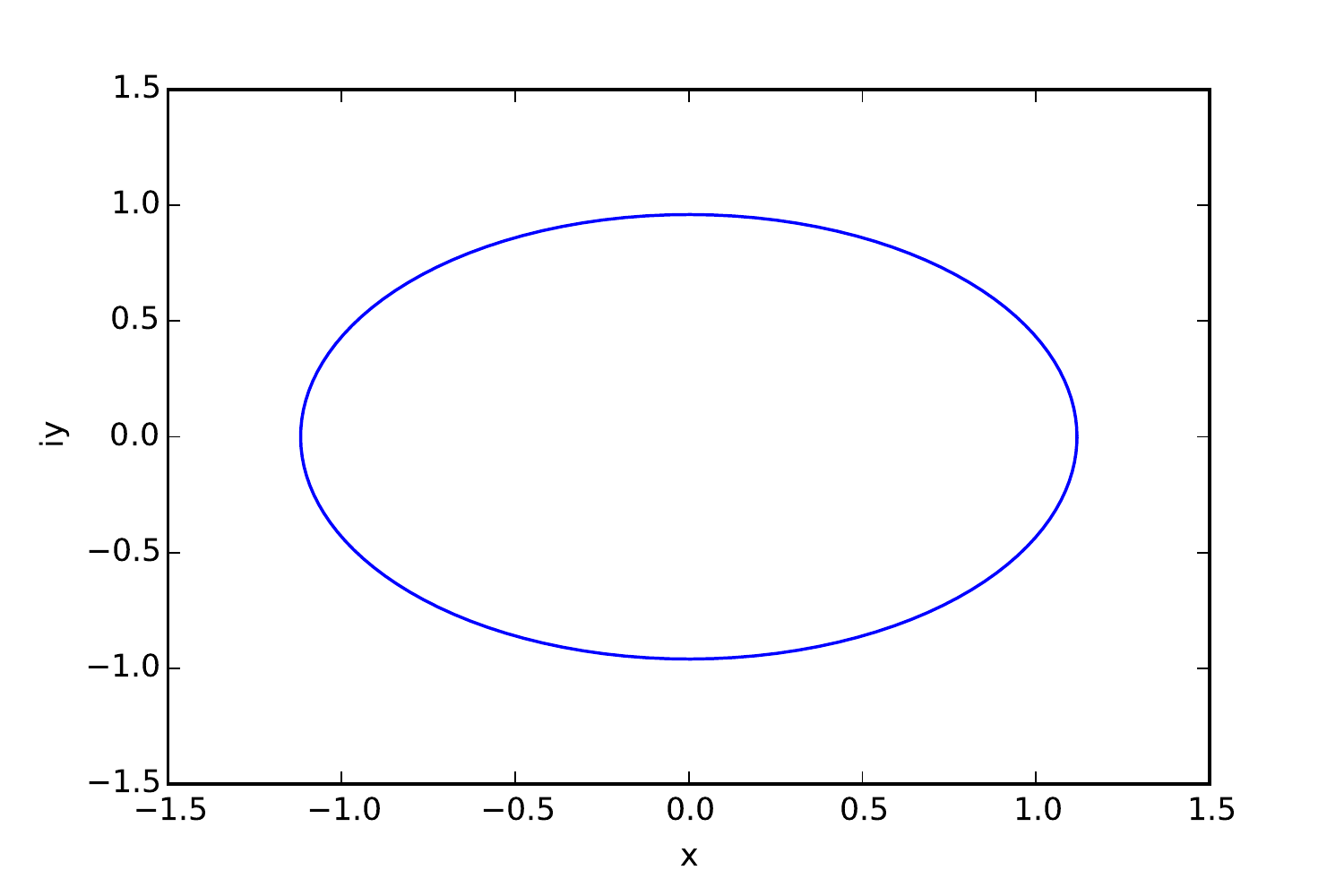}
\end{center}
\caption{
(Color online) 
$\nu$ contour for $c_1=1, \ c_0=0.5 $.}
\label{contour_CW}
\end{figure}

\begin{figure}
\begin{center}
\includegraphics[width=0.48\textwidth]{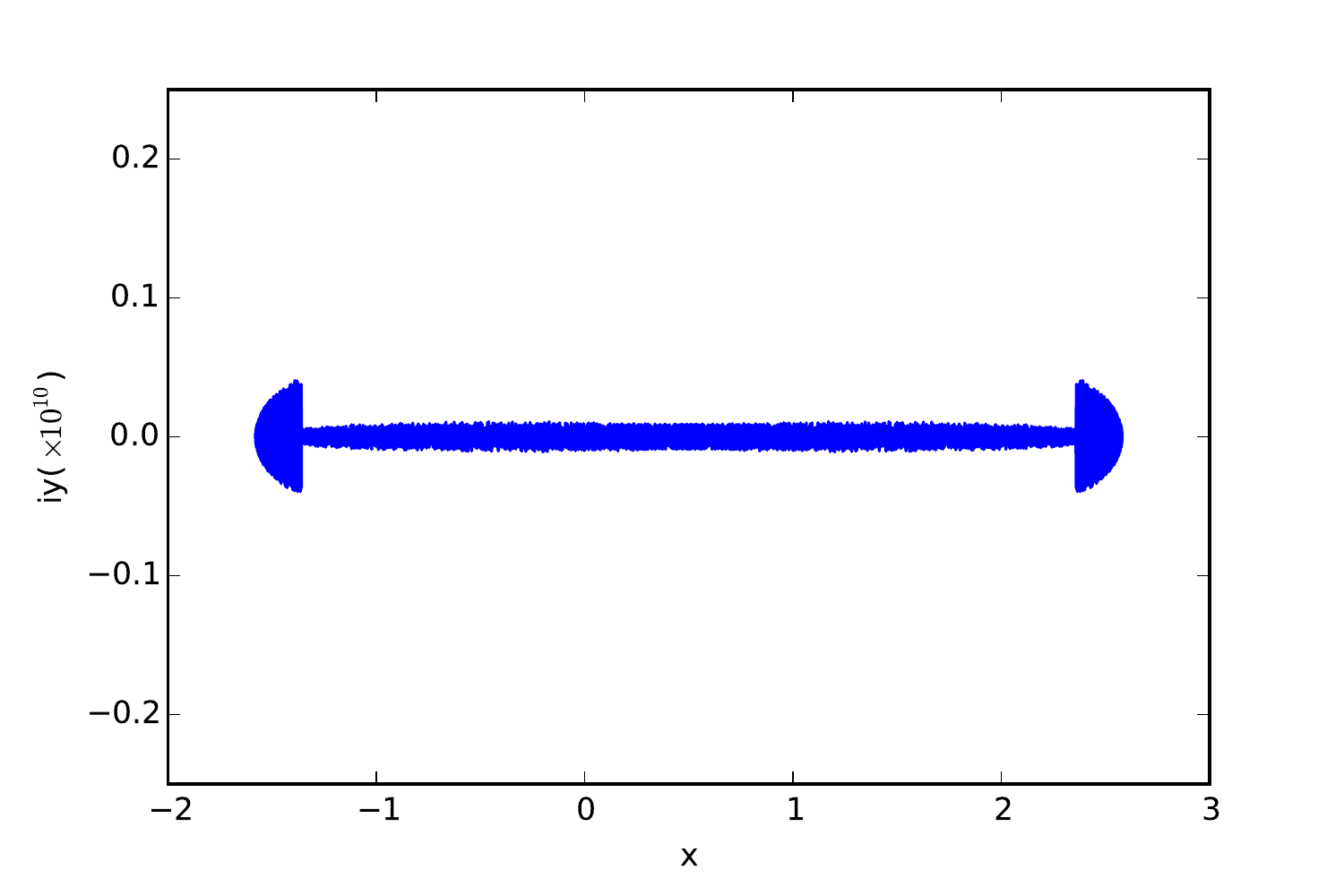}
\end{center}
\caption{
(Color online) 
Mapping for $\nu_1$ contour, $c_1=1, \ c_0=0.5$. Mapping for $\nu_2$ looks similar.}
\label{mapping_CW}
\end{figure}

Following CW, the JT for model (\ref{model-e}) is 
\begin{equation}
J_{c_1,c_0}(s) = c_1s+c_0-\log\frac{s-\frac{1}{2}}{s+\frac{1}{2}}
\label{joukowsky:CW}
\end{equation}
where $s$ is a complex variable. Note that the transformation now contains two parameters $c_0$ and $c_1$ to include the two supports for the soft-edges given by $[a,b]$ where both $a$ and $b$ are real numbers such that $a<b$. The JT is analytic in $\mathbb{C}\backslash[-\frac{1}{2},\frac{1}{2}]$ and has critical points on real line at $S_a=-\sqrt{\frac{1}{4}+\frac{1}{c_1}}$ and $S_b=\sqrt{\frac{1}{4}+\frac{1}{c_1}}$ which are mapped to points $a=J_{c_1,c_0}(S_a)$ and $b=J_{c_1,c_0}(S_b)$ respectively. There also exist points in the complex plane which are mapped to real line between $a$ and $b$ by $J_{c_1,c_0}(s)$. The equation of locus of such points is given by
\begin{equation}
x^2=\frac{1}{4}+\frac{y}{\tan(c_1y)}-y^2.
\label{locus_hard-edge:CW}
\end{equation}
Eq.~(\ref{locus_hard-edge:CW}) above forms a closed contour $\nu$ in complex plane which is symmetric about x-axis. We denote the two symmetric parts as curves $\nu_1$ and $\nu_2$ which are complex conjugates of each other, such that $\nu_1$ in the upper-half plane from $S_a$ to $S_b$, and a curve $\nu_2$ in the lower-half plane from $S_b$ to $S_a$. We note that $J_{c_1, c_0}$ maps the exterior of $\nu$ to $\compC$, and the interior of $\nu$, except for the interval $[-\frac{1}{2}, \frac{1}{2}]$, to the strip $\mathbb{S} := \{ x + iy \mid x \in \mathbb{R}, -\pi < y < \pi \}$.
Figure \ref{contour_CW} and Figure \ref{mapping_CW} show contour $\nu$ for $c_1=1, c_0=0.5$ and its mapping respectively.

\begin{figure}
\begin{center}
\includegraphics[width=0.48\textwidth]{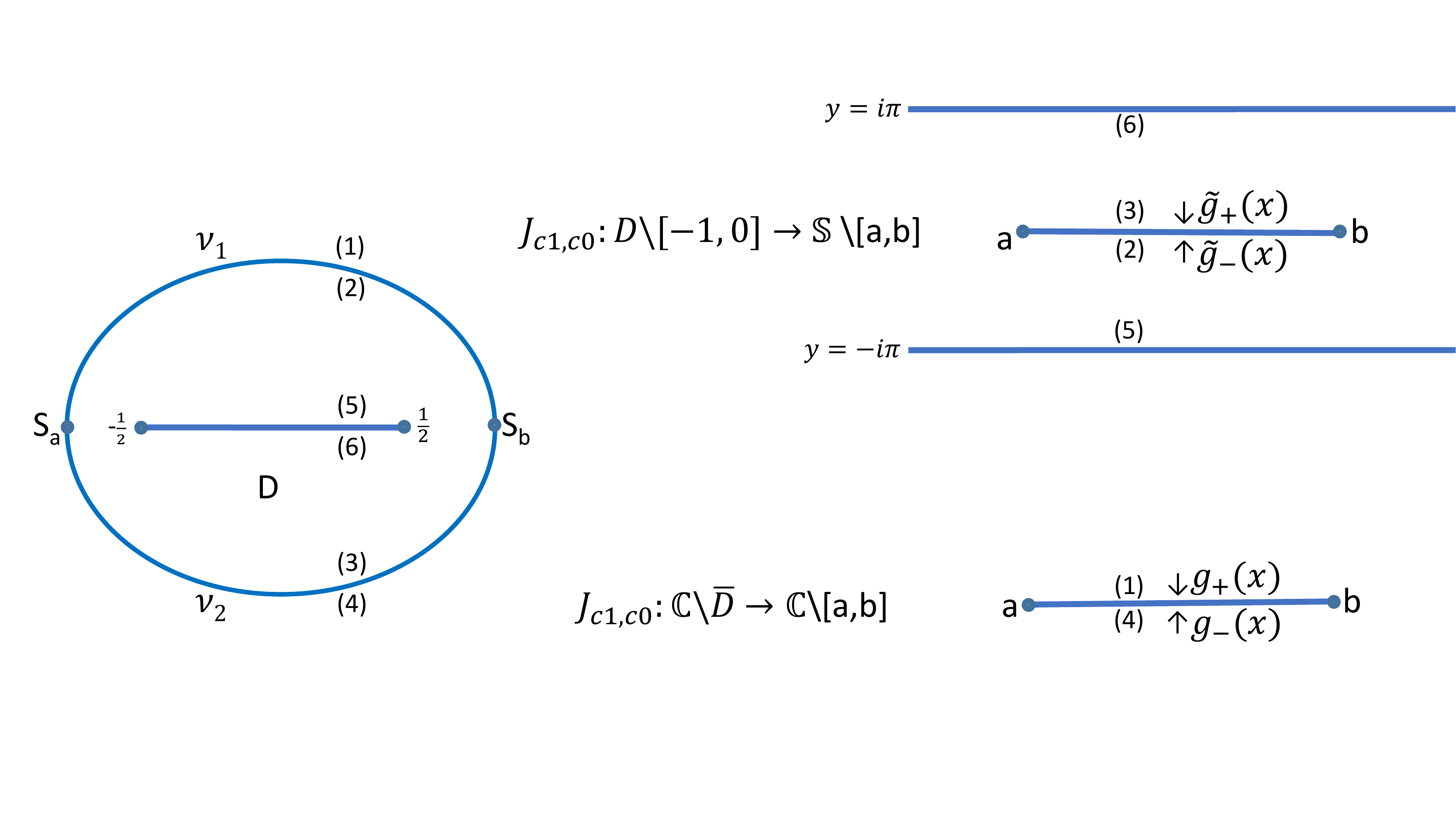}
\end{center}
\caption{
(Color online) 
Schematic Figure for mapping of JT, following CW.}
\label{mapCW}
\end{figure}
Figure \ref{mapCW} shows schematically the mapping of all points on contour $\nu$ and all the regions in complex plane respectively by the JT $J_{c_1,c_0}(s)$.
 All points except the branch cut $[-\frac{1}{2},\frac{1}{2}]$ inside region $D$ bounded by contour $\nu$ are mapped to complex region $\mathbb{S} \backslash[a,b]$. All the points outside region $D$ are mapped to a different complex region $\mathbb{C}\backslash[a,b]$.
 
We follow the method developed in Sections \ref{sec:3} and \ref{sec:4} to obtain an integral equation for the function $f(J_{c_1,c_0}(s))$.
The $g$-functions of Eq.~(\ref{complex_transforms}) are now replaced by

\begin{equation}
\begin{aligned}
  g_e(z) \equiv {}& \int_a^b \log (z-x)d\mu(x), && z \in \mathbb{C}\backslash(-\infty,b]; \\
  \tilde{g}_e(z) \equiv {}& \int_a^b \log (e^{z}-e^{x})d\mu(x), && z \in \mathbb{S}\backslash(-\infty,b].
\end{aligned}
\label{complex_transforms:CW}
\end{equation}
Here $(g_e,\tilde{g}_e)$ are analytic in $(\mathbb{C}\backslash(-\infty,b),\mathbb{S}\backslash(-\infty,b))$ respectively so that the logarithms are well defined. We note that in the $\gamma = 1$ case, $(g_e, \tilde{g}_e)$ satisfies a vector-valued RH problem that is similar to the RH problem for $(g, \tilde{g})$ given in Section \ref{sec:2}. Please see CW for detail. Let $g_{e+},g_{e-}$ and $\tilde{g}_{e+},\tilde{g}_{e-}$ denote boundary values of $g_e$ and $\tilde{g}_e$ when approaching $[-\infty,b]$ respectively from above $(+)$ and below $(-)$. The $M$-functions of Eq.~(\ref{M_def}) are replaced by 
\begin{equation}
M_e(s)\equiv
\begin{cases}
  G_e(J_{c_1,c_0}(s)), & \text{for } s\in\mathbb{C}\backslash \bar{D}, \\
  \tilde{G}_e(J_{c_1,c_0}(s)), & \text{for } s\in D\backslash[-\frac{1}{2},\frac{1}{2}],
\end{cases}
\label{M_def:CW}    
\end{equation}
where as before, $G_e(s)\equiv g_e^{\prime}(s)$ and $\tilde{G}_e(s)\equiv \tilde{g}_e^{\prime}(s)$.
The EL Eq.~(\ref{sum-difference}) remains the same, except that $J$ is now a function of two parameters $c_0$ and $c_1$. The function $f_e(J_{c_1,c_0})(s)$ is now defined as
\begin{equation}
f_e(J_{c_1,c_0})(s))\equiv M_{e+}(s_1)+M_{e-}(s_1)=M_{e-}(s_2)+M_{e+}(s_2)
\end{equation}
with solution to $M_e(s)$ as,
\begin{equation}
M_e(s) =
\begin{cases}
  \frac{-1}{2\pi i}\oint_{\nu}\frac{f_e(J_{c_1,c_0})(\xi))}{\xi -s}\; d\xi, & s\in \mathbb{C}\backslash \bar{D}, \\
  \frac{1}{2\pi i}\oint_{\nu}\frac{f_e(J_{c_1,c_0})(\xi))}{\xi -s}\; d\xi, & s\in D\backslash [-\frac{1}{2},\frac{1}{2}].
\end{cases}
\label{M_def_contr:CW}
\end{equation} 
As in Eq.~(\ref{inversemap}) before, we define the inverse mapping,
\begin{equation}
s_e=J_{c_1,c_0}^{-1}(x)=h_e(x).
\end{equation}
Note that for both $M_{e+}(s_1)$ and $M_{e-}(s_2)$ in Eq.~(\ref{M_def:CW}), the function is the limit of $M(s)$ as $s \in \mathbb{C} \setminus \bar{D}$ approaches $s_1$ or $s_2$ on contour $\nu$ from outside. Hence we used first identity in Eq.~(\ref{M_def_contr:CW}). Let $(s_1)_{e+} = h_e(y) \ ; \ (s_2)_{e-} = \bar{h}_e(y) \ ; \ s_{1e} = h_e(x) \ \text{and} \ s_{2e} = \bar{h}_e(x) $ where the bar denotes complex conjugate. In terms of the inverse mapping, the integral equation for $f_e$ now has the form,
\begin{equation}
\label{f_integral_eqn:CW}
f_e(y;\gamma)=\frac{V^{\prime}(y)}{\gamma} -\frac{1-\gamma}{\gamma 2\pi}\int_a^b f_e(x;\gamma)\phi_e(x,y)dx
\end{equation}
where
\begin{equation}
\phi_e(x,y)=\Im\bigg[ \left( \frac{1}{h_e(y) - \overline{h}_e(x)} + \frac{1}{\overline{h}_e(y) - \overline{h}_e(x)} \right) \overline{h}_e^{\prime}(x) \bigg].
\end{equation}
As given in CW, the JT parameters $c_1,c_0$ satisfy the following equations,
\begin{equation}
\begin{gathered}
  \frac{1}{2\pi i}{\displaystyle \oint_{\nu}^{}}U_{e ; c_1,c_0}(s)ds=\frac{1}{c_1}, \quad
  \frac{1}{2\pi i}{\displaystyle \oint_{\nu}^{}}\frac{U_{e;  c_1,c_0}(s)}{s-\frac{1}{2}}ds=1, \\
  U_{c_1,c_0}(s)=f_e(J_{c_1,c_0}(s)).
\end{gathered}
\label{c1c2}
\end{equation}
We solve the above integral equation  (Eq.~\ref{f_integral_eqn:CW})  for $f_e(y; \gamma)$ and Eq.~(\ref{c1c2}) for $c_1, c_0$  numerically self-consistently. Using the definition for $f_e(x;\gamma)$ we further find the new effective potential $V_{\eff}(x;\gamma)$ which is related to $f_e(x;\gamma)$ by 
\begin{equation}
V'_{\eff}(x;\gamma)=f_e(x;\gamma).
\label{V-effective:CW}
\end{equation} 
The  corresponding density is computed using the formula from CW,
\begin{equation}
\sigma_{e}(y)=\frac{-1}{2\pi i}[M_{e+}(s_{e1})-M_{e-}(s_{e2})].
\end{equation}

Substituting for $M_{e+}(s_{e1})$ and $M_{e-}(s_{e2})$, the expression for density becomes
  
\begin{equation}
\label{density_soft_edge:CW}
\sigma_{e}(y;\gamma)=\frac{-1}{2{\pi}^2}\int_b^a f_e(x;\gamma)\chi_e(x,y) dx ,
\end{equation}

where

\begin{equation}
\chi_e(x,y)=\Re\bigg[ \bigg( \frac{1}{\overline{h}_e(y) - h_e(x)}-\frac{1}{h_e(y) - h_e(x)} \bigg)h_e^{\prime}(x)\bigg].
\end{equation}

\section*{References}


\end{document}